\documentstyle[aps,pra,multicol,epsfig]{revtex}
\begin{document}
\draft

\title{The generalized contact process with $n$ absorbing states}
\author{Jef Hooyberghs$^{1,2}$, Enrico Carlon$^3$ and Carlo 
Vanderzande$^{1,4}$}
\address{
$^1$ Departement WNI, Limburgs Universitair Centrum, 3590 Diepenbeek, Belgium
\\
$^{2}$ Aspirant Fonds voor Wetenschappelijk Onderzoek, Vlaanderen \\
$^3$ INFM, Dipartimento di Fisica, Universit\`a di Padova,
I-35131 Padova, Italy\\
$^{4}$ Instituut voor Theoretische Fysica, Celestijnenlaan 200D, 3001
Heverlee, Belgium}
\date{\today}
\maketitle

\begin{abstract}
We investigate the critical properties of a one dimensional stochastic lattice 
model with $n$ (permutation symmetric) absorbing states. We analyze the cases 
with $n \leq 4$ by means of the non-hermitian density matrix renormalization 
group. For $n=1$ and $n=2$ we find that the model is respectively in the 
directed percolation and parity conserving universality class, consistent with 
previous studies. For $n=3$ and $n=4$, the model is in the active phase in the 
whole parameter space and the critical point is shifted to the limit of one 
infinite reaction rate. 
We show that in this limit the dynamics of the model can be mapped onto that 
of a zero temperature $n$-state Potts model.
On the basis of our numerical and analytical results we conjecture that 
the model is in the same universality class for all $n \geq 3$ with exponents 
$z = \nu_\|/\nu_\perp = 2$, $\nu_\perp = 1$ and $\beta = 1$. These exponents 
coincide with those of the multispecies (bosonic) branching annihilating 
random walks. For $n=3$ we also show that, upon breaking the symmetry to a 
lower one ($Z_2$), one gets a transition either in the directed percolation, 
or in the parity conserving class, depending on the choice of parameters.
\end{abstract}


\begin{multicols}{2} \narrowtext

\section{Introduction}

The study of systems out of thermal equilibrium has attracted great
attention in recent years. As their equilibrium counterparts,
these systems may display continuous phase transitions characterized by a
set of critical exponents \cite{outeq}. In particular, much interest
exists in
transitions from a
fluctuating active state towards an absorbing state, i.e. a configuration
where the dynamics is frozen.

A prototype of a 1D lattice model with a transition into an absorbing state is
the contact process \cite{outeq,CP}, in which each lattice site can be either 
empty ($\emptyset$) or occupied by a particle ($A$) with the reactions $A 
\to2A$, $A \to\emptyset$. Depending on the relative rates of these processes,
the stationary state is empty (when $A \to\emptyset$ dominates), or is occupied
by a finite density of particles (if the reaction $A \to2A$ dominates). The 
contact process therefore displays a non-equilibrium phase transition which is 
known to belong to the directed percolation (DP) universality class. The empty 
lattice ($\emptyset \emptyset\ldots\emptyset$) is the absorbing state.

A wide range of models with transitions into absorbing states was found to
belong to the DP universality class. As typical examples we quote the
branching-annihilating random walks (BARW) with an odd number of offsprings
\cite{BARWo}, the Domany-Kinzel model \cite{DK} and the pair contact process
\cite{PCP}. The DP class therefore appears to be extremely robust and quite
common, but it is certainly not the only possible one.

Another universality class that has by now been firmly established is the
so-called parity conserving (PC) class \cite{outeq}. A prototype model 
in this class is the BARW with an even number of offsprings \cite{BARWe} in 
which particles can diffuse and undergo the reactions $2 A \to\emptyset$, $A 
\to(m+1)A$, with $m$ an even integer. In that model, the particle conservation
modulo two is believed to be the reason for the system to show non-DP
critical behavior. More recently it became clear that the parity conservation,
at least at the microscopic level,
is not a necessary condition for a PC transition to occur
\cite{Park95,Hinr97}. Hinrichsen \cite{Hinr97} provided an example of this by
introducing a one dimensional model where each lattice site can be occupied 
by at most one particle or can be in any of $n$ inactive states ($\emptyset_{1}$,
$\emptyset_{2}$ \ldots$\emptyset_{n}$). The reactions are:
\begin{eqnarray}
AA   \rightarrow  A \emptyset_k,\emptyset_k A
&&
\ \ \ \ \ \ \
\mathrm{with\ rate\ \lambda/n}\label{reac_1}\\
A\emptyset_k ,\emptyset_k A  \rightarrow \emptyset_k \emptyset_k
&& 
\ \ \ \ \ \ \
\mathrm{with\ rate\ \mu_k}\label{reac_2}\\
A\emptyset_k ,\emptyset_k A  \rightarrow AA
&&
\ \ \ \ \ \ \
\mathrm{with\ rate\ 1}\label{reac_3}\\
\emptyset_k \emptyset_l \rightarrow A \emptyset_l,\,\,\emptyset_k A
&&
\ \ \ \ \ \ \
{(k\neq l)\,\,
\mathrm{with\,\,\,rate\,\,\,1}
}\label{reac_4}
\end{eqnarray}
We refer to the model defined by the reactions (\ref{reac_1}-\ref{reac_4}) as 
to the generalized contact process (GCP).
The original contact process \cite{CP}, corresponds to the case $n=1$, in 
which the reaction (\ref{reac_4}) is obviously absent. Notice that the 
reaction (\ref{reac_4}) in the case $n \geq2$ ensures that configurations 
as $(\emptyset_{i}
\emptyset_{i} \ldots\emptyset_{i} \emptyset_{i} \emptyset_{j} \emptyset_{j}
\ldots\emptyset_{j} \emptyset_{j})$, with $i \neq j$ are not absorbing. Such
configurations do evolve in time until the different domains coarsen and one
of the $n$ absorbing states $(\emptyset_{1} \emptyset_{1} \ldots\emptyset
_{1})$, $(\emptyset_{2} \emptyset_{2} \ldots\emptyset_{2})$, \ldots
$(\emptyset_{n} \emptyset_{n} \ldots\emptyset_{n})$ is reached.

For the GCP with $n=2$, it was found from simulations \cite{Hinr97}
that the transition falls in the PC class if $\mu_{1} = \mu_{2}$, while if 
the symmetry between the two absorbing states was broken ($\mu_{1} \neq
\mu_{2}$) a DP transition was recovered \cite{Hinr97}.

One aim of this paper is to investigate the case $n \geq3$, which has not been 
studied so far. Our results are obtained by a numerical investigation based on 
density matrix renormalization group (DMRG) techniques \cite{DMRGbook} for the 
case $n=3$ and $n=4$ and an exact solution in the limit $\mu \to\infty$. 
On the basis of these results we are led to conjecture that for
$n \geq 3$ the model is always in the same universality class, which
coincides with that of multispecies branching and annihilating random
walks \cite{Card98}.

There are several reasons which make the GCP interesting. 
Firstly, it contains the two mayor universality
classes for transitions into absorbing states, namely DP ($n=1$) and PC ($n=2$).
Secondly, for $n > 1$ is it interesting to study the effect of breaking the
permutational symmetry of the model and for $n > 2$ the symmetry can be broken
in different ways (see below). Finally, it is also interesting to investigate
the performance of the DMRG algorithm for a system with several absorbing
configurations and with several states per site, a situation which is
definitely more complicated than what was considered so far \cite{dmrg,2a3a}.
We recall that the application of the DMRG to non-equilibrium processes is 
rather recent and its power/limitations have not been fully investigated yet, 
therefore the GCP represents another important testing ground for this 
purpose.

This paper is organized as follows: in Sec. \ref{sec:pd} we present the
numerical results for the cases $n=1,2,3$ and $4$. In Sec. \ref{sec:exact}
we show that in the limit $\mu \to \infty$ the dynamics of the model can be 
mapped onto that of the zero temperature $n$-state Potts model, which allows 
the determination of one critical exponent.
Next, we present in Sec. \ref{sec:conjecture} a conjecture for the critical 
behavior of the model for $n \geq 3$ based on the numerical and exact results.
Finally, in Sec. \ref{sec:breaking} we analyse the effect of breaking the
symmetry in the $n=3$ case, while Sec. \ref{sec:concl} concludes the paper.

\section{DMRG study of the model}

\label{sec:pd}

As a starting point for our analysis we use the quantum form of the master
equation \cite{kwantum,schutz} to describe the evolution of 
the stochastic processes in continuous
time:
\begin{equation}
\frac{d|P(t)\rangle}{dt}=-H|P(t)\rangle \label{f1}
\end{equation}
where $|P(t)\rangle$ is a state vector whose elements are the probabilities of
finding the system in a certain configuration, and the entries of the matrix
$H$ are the transition probabilities per unit of time between different
configurations. As in quantum mechanics we will call $H$ the Hamiltonian
of the system. However, in the most general case, as in the GCP, the
matrix $H$ is non-hermitian, therefore one should distinguish between
right and left eigenvectors, which are now not related by transposition. As a
second consequence, the eigenvalues could be complex, but $H$ always has
at least one eigenvalue that is zero, and the real part of the non vanishing 
eigenvalues is strictly positive. Since we are interested in the stationary 
behavior of the system and the relaxation towards it, we will determine the low
lying spectrum of the Hamiltonian, i.e. the part of the spectrum with the
smallest real part. In sections II-IV we will consider the system to be
symmetric in the ground states: $\mu_{1}=\mu_{2}=\ldots=\mu$.

First of all, the conservation of probability always ensures the existence of
a trivial left eigenvector with zero eigenvalue $\langle0|\equiv\sum_{\sigma
}\langle\sigma|$, where the sum is extended over all possible configurations
$\left\langle \sigma\right|$. Besides this left ground state, the GCP
has $n$ trivial right ground states: the $n$ absorbing configurations
$|\psi_{k}^{0}\rangle\equiv|\emptyset_{k}\emptyset_{k}\ldots\emptyset
_{k}\rangle$, with $k=1,2\ldots n$. To study the rest of the low lying
spectrum we apply the DMRG in combination with finite size scaling. The DMRG
is an iterative algorithm through which one constructs approximate eigenvalues
and eigenvectors of $H$ for long chains. At each iteration the lattice
size is increased and the configurational space is truncated efficiently, so
that one considers, instead of the exact operator $H$, effective
matrices of reduced dimensions, which can be handled numerically. The accuracy
obtained for the dominant eigenvalues and eigenvectors is often very good.
Although originally invented for hermitian problems, the DMRG also works in
non-hermitian cases, as it has been shown in several examples \cite{dmrg,2a3a}.

\subsection{The cases $n=1$ and $n=2$}

For $n>1$ the GCP has more than one absorbing state and 
therefore more than two
states per site. This, 
together with the fact that the Hamiltonian is non-hermitian,
makes a DMRG study of the model technically difficult. Consequently it is of
great importance to have a way to check the results of our method. Until now
the GCP has only been studied by means of Monte Carlo simulations, 
for $n=1,2$ \cite{Hinr97}. Therefore, this subsection is
restricted to these two cases. We will explain how we applied the DMRG, give
some details on the finite size scaling for $n=2$ and compare our results with
\cite{Hinr97}.

A first quantity which can be calculated by the DMRG is the gap $\Gamma$
between (the real part of) the eigenvalue of the first excited state of the
Hamiltonian $H$ and the absorbing ground states $\left|  \psi_{k}
^{0}\right\rangle $. $\Gamma$ is the inverse relaxation time and enables us to
determine the critical region and the dynamical critical exponent
$z=\nu_{\parallel}/\nu_{\perp}$. A direct implementation of the DMRG is
however very unpractical since for $n>1$ we have a degenerate ground state,
and the DMRG is known to work best for gapped systems. We alleviated this
problem by adding at the two boundary sites the following reactions

\begin{equation}
\emptyset_{k}\rightarrow\emptyset_{1}\,\,\,\,\,\,\,\,{(k\neq
1)\,\,\mathrm{with\,\,\,rate\,\,\,p}}\label{reac_5}
\end{equation}
(Recall that it is customary to use open boundary conditions in DMRG
\cite{DMRGbook}). 
With (\ref{reac_5}), only $|\psi_{1}^{0}\rangle$ is left as a ground state. 
The bulk critical behavior is however expected to be unchanged. Next we
performed the transformation \cite{nota}:
\begin{equation}
H^{\prime}(\Delta)\equiv H+\Delta|\psi_{1}^{0}\rangle
\langle0|\label{shift}
\end{equation}
with $\Delta>0$ and where $H$ is constructed from the reactions
(\ref{reac_1}-\ref{reac_4}) and (\ref{reac_5}). $H^{\prime}$ is no
longer a stochastic Hamiltonian, the zero eigenvalue is shifted to $\Delta$,
but the rest of its spectrum is exactly the same as that of $H$. This
can easily be checked by looking at the eigenvalues of the left eigenvectors
which are the same for both matrices. Therefore the calculation of the gap of
the original Hamiltonian with an $n$-times degenerate ground state is reduced 
to the calculation of the lowest eigenvalue of $H^{\prime}$ (obviously provided
$\Delta$ is bigger than the gap of $H$). This strategy could be generalized to 
other systems with several absorbing states, provided one finds an appropriate 
boundary reaction which ``eliminates" zero eigenstates, as done by 
(\ref{reac_5}) for the GCP.

\begin{figure}[b]
\centerline{
\psfig{file=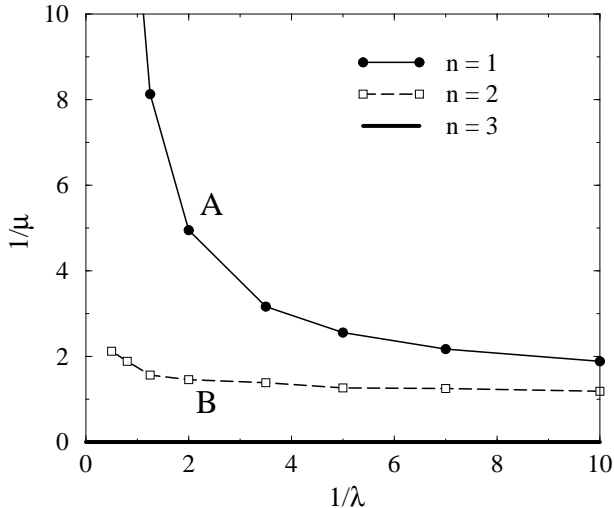,height=7cm}}
\vskip 0.2truecm   
\caption{Phase diagram of the generalized contact process for 
$\protect\mu _{1} = \protect\mu
_{2} = \ldots \protect\mu _{n} = \protect\mu $, as obtained by non-hermitian
DMRG techniques, for $n=1$ (filled circles), $n=2$ (empty squares) and $n=3$
(thick solid line). For $n=1$ and $n=2$, the region above the curves is the
active phase, while the region below is inactive. 
In the limit $\protect\lambda \to \infty$ the critical line for $n=2$ 
approaches a finite value of $\mu$, while that for $n=1$ diverges towards
$\mu \to 0$. In the other limit
$\protect\lambda \to 0$ both lines merge into a common special point (see
Ref. \protect\cite{Hinr97}).
For $n=3$, the model is active in the whole parameter space except along the 
critical line $1/\protect\mu =0$.}
\label{FIG01}
\end{figure}

Before turning to the finite size analysis of the calculated gap, let us first
present the phase diagram for $n=1$ and $n=2$. Figure
\ref{FIG01} shows these diagrams, obtained by the DMRG method and standard
finite size scaling techniques. The region above the lines denotes the active
state, where the system has a finite stationary density of particles, while
below these lines the stationary density is zero and the system is in one of
the absorbing states. We find (see below) that the critical exponents are the
same all along the
critical line and are those of the DP and PC universality class for $n=1$ and
$n=2$ respectively. We notice that the active region increases from $n=1$ to
$n=2$. The location of the critical lines agrees well with Monte Carlo
simulations by Hinrichsen \cite{Hinr97}.

It is worth while at this point to present in some more detail the finite size
scaling analysis employed for $n=2$, in order to clarify the differences with
the higher $n$ case. For each choice of the rates $(\lambda,\mu)$ we can
calculate the gap $\Gamma_{L}$ of the matrix $H$ for a system of length $L$. 
This gap equals the lowest eigenvalue of $H^{\prime}$ defined in 
Eq. (\ref{shift}).
As a function of L, $\Gamma_{L}$ has a different scaling behavior in the
three regions of the phase diagram. In the active phase the gap should scale
as $\Gamma_{L}\sim\exp{(-aL)}$, since asymptotically in L the absorbing states
are degenerate with a state having a finite density of particles. At the
critical line we have $\Gamma_{L}\sim L^{-z}$, with $z=\nu_{\Vert}/\nu_{\perp}$
the dynamical exponent. 
For $n=2$ the system is in the PC universality class and the whole inactive 
phase is known to be algebraic: the gap decays as $\Gamma_{L}\sim L^{-2}$. 
Notice that the latter behavior is different in DP models where the gap is
finite, i.e. $\Gamma_{L} \sim \Gamma_0 >0 $, in the inactive phase.

The finite size scaling behavior of $\Gamma_{L}$ can best be monitored by 
plotting the discrete logarithmic derivative of the gap: 
$Y_{L}\equiv\ln(\Gamma_{L+1}/\Gamma_{L-1})/\ln[(L+1)/(L-1)]$. In the
active phase the scaling form $\Gamma_{L}\sim\exp{(-aL)}$ implies $Y_{L}
\sim -aL$ (with $a$ a positive constant), while $Y_{L} \sim -z$ at the critical 
line.

\begin{figure}[b]
\centerline{
\psfig{file=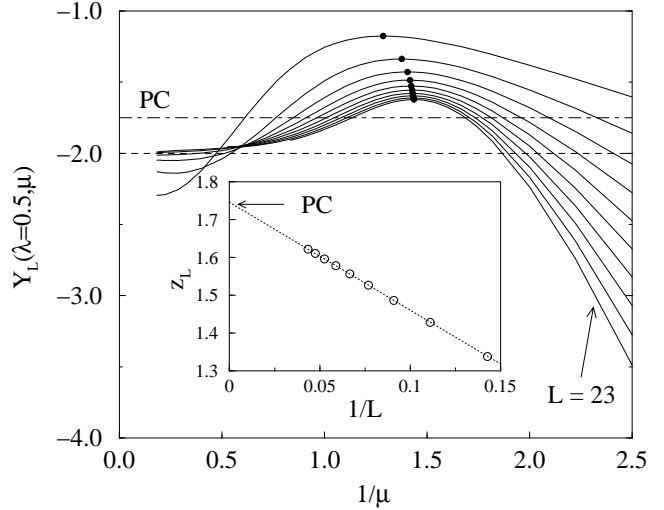,height=7cm}}
\vskip 0.2truecm 
\caption{Plot of $Y_L$ for the GCP with $n=2$ as function of $\protect
\mu$ and for $\protect\lambda = 0.5$. Inset: plot of $z_L$ as function of
the inverse system length $1/L$. The horizontal lines indicate the values 2
and the known value of the exponent $z$ for the PC universality class ($
z=1.74$).}
\label{FIG02}
\end{figure}

Figure \ref{FIG02} shows a plot of $Y_{L}$ for $n=2$, $\lambda=0.5$ as a
function of the parameter $\mu$ for $L=5,7\ldots23$ (in the present case the
calculation of the gap was extended up to $L=24$). In the right part of the
figure one distinguishes the scaling for the active phase, while in the left
part (large $\mu$) one sees that $Y_{L}$ flattens out and approaches the value
of $-2$, as expected for the algebraic inactive phase of PC models. The maxima
of $Y_{L}$, marked with a filled circle in Fig. \ref{FIG02}, identify the
boundary between the active and inactive phase and can be used as critical
point estimates. From an $L\rightarrow\infty$ extrapolation we find $\mu_{c}
=0.69(1)$ for our choice of $\lambda=0.5$, which determines point B of the
critical line in Fig. \ref{FIG01}. In the inset of Fig. \ref{FIG02} we plot 
$z_{L}=-\max_{\mu }Y_{L}(\mu,\lambda=0.5)$, as function of $1/L$. 
Extrapolating $z_{L}$ in the limit $L\rightarrow\infty$ we find 
$z_{L} \rightarrow 1.747$, in accurate agreement with the corresponding 
exponent known for the PC class $z=\nu _{\Vert}/\nu_{\perp}=1.74(1)$ 
\cite{nota1}.

In this way, the DMRG enables us to determine the critical region and the
exponent $z$. As we will show in the following subsection, using different
boundary conditions we can also find estimates for other exponents. In Table
\ref{TABLE01} we show our estimates of the critical exponents $z$ and
$\beta/\nu_{\perp}$, in the points A and B of Fig. \ref{FIG01}. We recover the
exponents of DP and PC as expected for $n=1$ and $n=2$ respectively,
indicating that the DMRG is performing well.

%
\vbox{
\begin{table}[tb]
\caption{Estimates for the critical exponents $z = \nu_\|/\nu_\perp$ and
$\beta/\nu_{\perp}$ calculated for $n = 1$ and $2$ absorbing states. We recall
that for DP $z = 1.5806$, $\beta/\nu_\perp= 0.25$ and for PC
$z = 1.75$, $\beta/\nu_\perp= 0.50$.
}
\label{TABLE01}
\begin{tabular}{ccccc}
$n$  & $\lambda$ & $\mu_{c}$ & $z=\nu_\|/\nu_\perp$ & $\beta/\nu_\perp$ \\
\hline
1 & 0.5 & 0.20(1) & 1.575(5) & 0.255(5) \\
2 & 0.5 & 0.69(1) & 1.747(5)  & 0.49(1)
\end{tabular}
\end{table}
}
%

\subsection{The case $n=3$}

\begin{figure}[b]
\centerline{
\psfig{file=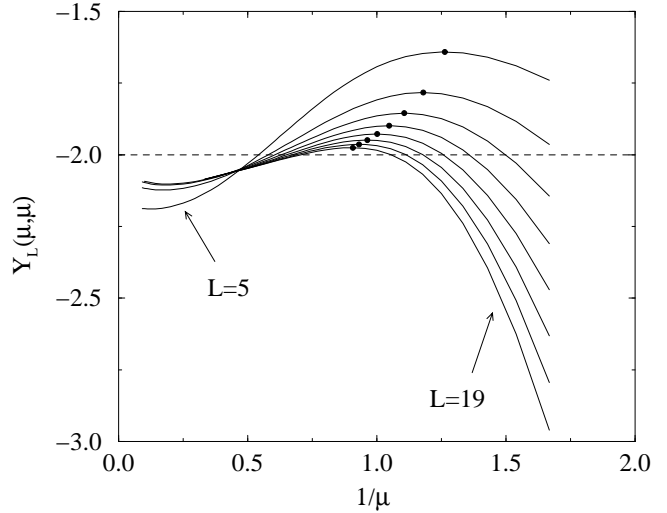,height=7cm}}
\vskip 0.2truecm 
\caption{Plot of $Y_{L}$ for the GCP with $n=3$ along the line $
\protect\lambda =\protect\mu $.}
\label{FIG03}
\end{figure}

Since the results of the DMRG are consistent with the Monte Carlo 
results
for $n=1$ and $n=2$, we can confidently use it to study also
the case $n=3$.
 Again we
start by an analysis of the gap: in Fig. \ref{FIG03} we plot the quantity
$Y_{L}$ along the line $\lambda=\mu$ for $L=7,9\ldots19$. As for $n=2$ we find
a clear evidence of an active phase for small $\mu$ where $Y_{L}\sim-aL$.
However, when we want to determine the critical point by localising the
maxima, we find that upon increasing $L$ the maxima shift towards larger values
of $\mu$. In the limit $L\rightarrow\infty$ we find $\mu_{\mathrm{max}
}\rightarrow\infty$, indicating that the system is always active and that the
critical point is shifted to $1/\mu=0$. This numerical evidence 
together with the arguments presented in the next section lead us
to expect that the model is
critical only if $1/\mu=0$, as indicated in Fig. \ref{FIG01}.
The estimate of the exponent $z$ at $\lambda=\mu\rightarrow\infty$ yields
$z=2.00(3)$. In the next section we will give an analytical treatment of the
case $1/\mu=0$ confirming this value of $z$.

To get access to more critical exponents, we introduce different
boundary conditions. We replace (\ref{reac_5}) with:
\begin{equation}
\emptyset_{k}\rightarrow A\,\,\,\,\mathrm{with\,\,rate\,\,p^{\prime}
} \ \ \ \ \ \ \forall\ \mbox{k}
\label{inject}
\end{equation}
through which particles are continuously injected at the boundary sites. Now
the states $|\psi_{k}^{0}\rangle$ are no longer absorbing configurations, and
the spectrum of the Hamiltonian is non-degenerate. There is a unique right
eigenvector $|\phi_{0}\rangle$, which for finite system lengths always has a
finite density of particles. In the active phase the gap 
$\Gamma^{\prime}_{L}$ is no longer
asymptotically degenerate, as in the previous case, but it remains finite,
indicating that the system relaxes exponentially fast towards 
$|\phi_{0}\rangle$. Summarizing, if we consider the model defined by the 
reactions (\ref{reac_1}-\ref{reac_4}) and the boundary term (\ref{inject}) we 
have a gap behaving as $\lim_{L\rightarrow\infty} \Gamma^\prime_{L} =
\Gamma_{0}>0$ in the active phase, while as before $\Gamma^\prime_{L}\sim 
L^{-z}$ at the critical point. This means we can use $\Gamma^\prime_{L}$ 
itself to discriminate between the active and critical domains.

\begin{figure}[b]
\centerline{
\psfig{file=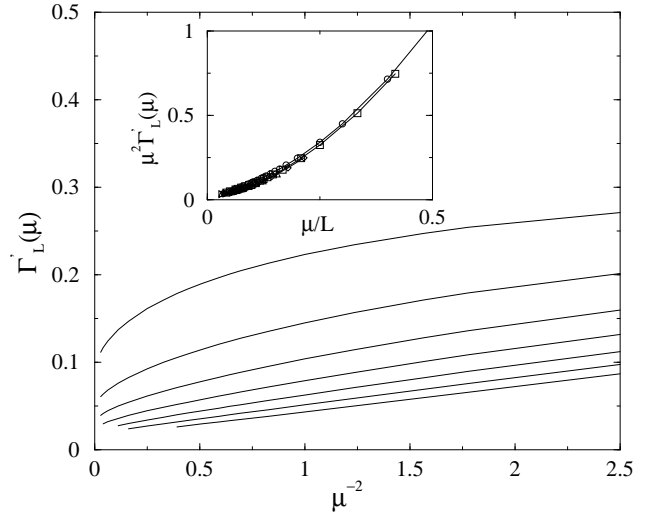,height=7cm}}
\vskip 0.2truecm 
\caption{Plot of the gap $\Gamma^\prime_{L}(\protect\mu)$,
calculated along the line $\protect\lambda =\protect\mu $ (with $p^\prime
=\protect\mu $ in reaction (\ref{inject})) as function of $\protect 
\mu ^{-2}$ for systems of lengths $L=6,8\ldots 18$. Inset: scaling collapse
of $\protect\mu ^{\protect\nu _{\Vert }}\Gamma^\prime_{L}(\protect\mu )$, 
plotted as function of $\protect\mu L^{-1/\protect\nu _{\perp }}$ for $L=10$,
$12$, $14$, $16$ and $18$ and where we used $\protect\nu _{\Vert }=2$ and 
$\protect\nu_{\perp }=1$.}
\label{FIG04}
\end{figure}

Figure \ref{FIG04} shows a plot of the gap $\Gamma^\prime_{L}$ along the line
$\lambda=\mu$. When we extrapolate $L\rightarrow\infty$, the gap remains
finite for every value of $\mu$, i.e. we find again that the system is active
throughout the whole parameter space.

These data also give the possibility to estimate the correlation length
exponent along the time direction: $\nu_{\parallel}$. Finite size scaling
around the critical point $1/\mu=0$ gives us the following scaling relation:
\begin{equation}
\Gamma^\prime_{L}(\lambda=\mu)=\mu^{-\nu_{\Vert}}
f\left(  \mu L^{-z/\nu_{\Vert}}\right)
\end{equation}
For $\mu$ fixed and $L\rightarrow\infty$, $\Gamma^\prime_{L}$ remains finite, 
so the scaling function should behave like $\lim_{x\rightarrow0^{+}}f(x)=
f_{0}$. As a consequence in the thermodynamic limit $L\rightarrow\infty$ the 
gap should vanish at the critical point as $\Gamma^\prime_{L}\sim
\mu^{-\nu_{\Vert}}$. The linear behavior of the gap in Fig. \ref{FIG04}, when 
plotted as a function of $\mu^{-2}$, already indicates that $\nu_{\Vert}=2$ 
and this results is also consistent with the scaling collapse reported in the 
inset, where we took $\nu_{\Vert}=2$ and $z=2$.

Finally using the same boundary term (\ref{inject}) it is also possible to
compute the critical exponent $\beta$ which describes the behavior of the 
particle density near the critical point. With the boundary condition 
(\ref{inject}) the stationary state of any finite system has a non-zero 
density of particles. We calculated in particular the average density of 
particles in the central site of the chain which we will denote as 
$\rho_{L}(\lambda,\mu)$.
This quantity, calculated along the line $\lambda=\mu$, is shown in
Fig. \ref{FIG03} for chains of various lengths (up to $L=24$) and plotted as
function of $1/\mu$. For large $\mu$ and $\lambda$ ($\mu \geq 2$) the DMRG 
does not perform so well and we had to restrict the calculation to $L=14$.

\begin{figure}[b]
\centerline{
\psfig{file=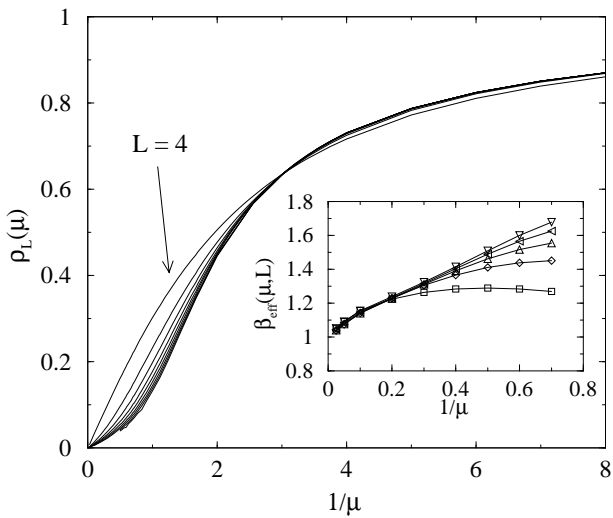,height=7cm}}
\vskip 0.2truecm 
\caption{Plot of the stationary particle density $\protect\rho $ for the
model with $n=3$ absorbing states along the line $\protect\mu =\protect
\lambda $ for various system sizes ($L=4,6,\ldots 24$). For large $\protect
\mu $ the calculation has been limited to $L=14$. Inset: plot of the
effective exponent $\protect\beta $ as function of $1/\protect\mu $ for $L=6$
(lower curve) up to $L=14$ (upper curve).}
\label{FIG05}
\end{figure}

As seen in the figure in
the limit $L\rightarrow\infty$ the stationary particle density vanishes
only for $\mu\rightarrow\infty$, again indicating that the system is always
active for any finite $\mu$. 
For large $\mu$ the order parameter is expected to decay as $\rho
\sim \mu^{-\beta}$ therefore the quantity:
\begin{equation}
\beta_{\mathrm{eff}}(\mu,L ) =-\frac{d\ln\rho_{L}(\lambda=\mu)}{d\ln\mu}
\end{equation}
converges to $\beta$ for $L \to \infty$ and $\mu \to \infty$.
The inset of Fig. \ref{FIG03} shows a plot of $\beta_{\mathrm{eff}}$ as
function of $1/\mu$ for chains of various system lengths. The extrapolated
result for $L=12$ and $L=14$ is $\beta=1.00(1)$.

In summary, we found that for $n=3$, the system is always active, except for 
$1/\mu=0$ where it is critical and we have exponents consistent with $z=2$, 
$\nu_{\parallel}=2$, $\beta=1$.

\subsection{The case $n=4$}

\begin{figure}[b]
\centerline{
\psfig{file=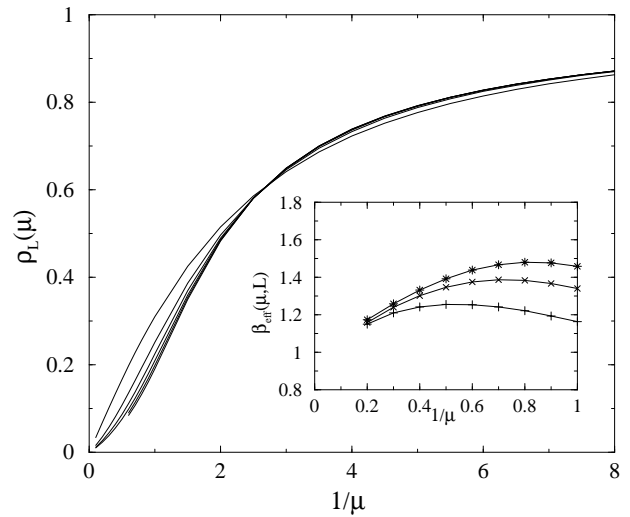,height=7cm}}
\vskip 0.2truecm 
\caption{Plot of the stationary particle density $\protect\rho $ for the
model with $n=4$ absorbing states along the line $\protect\mu =\protect
\lambda $ for various system sizes ($L=4,6,\ldots 14$). For large $\protect
\mu $ the calculation has been extended up to $L=10$. Inset: plot of the
effective exponent $\protect\beta $ as function of $1/\protect\mu $ for $L=6$
(lower curve) up to $L=10$ (upper curve).}
\label{FIG06}
\end{figure}

To conclude our numerical calculations for the case of symmetrical ground 
states, we studied the system with four inactive states per site. Here we 
restricted ourselves to a calculation of the density $\rho_{L}(\lambda,\mu)$ 
with boundary condition (\ref{inject}). Our results are shown in figure 
\ref{FIG06}.

From these data we find, as was the case for $n=3$ that the system is 
active for any finite value of $\mu$, and that the critical exponent 
$\beta=0.98(4)$.

\section{Fast rate expansion for $\mu$ or $\lambda \to \infty$}
\label{sec:exact}

In this section we show how the dynamics of the model can be simplified 
in the limit where one of the rates of the decay processes (\ref{reac_1})
or (\ref{reac_2}) goes to infinity. For $\mu \to \infty$ the resulting 
effective dynamics coincides with that of a zero temperature $n$-state
Potts model. This leads to a determination of the exponent $z$.
 
Intuitively, the argument goes as follows. For $\mu \to \infty$, all particles 
in the system will disappear very quickly, and in any finite system one will 
soon be in a configuration containing only the $n$ empty sites $\emptyset_1$, 
$\emptyset_2$ and $\emptyset_n$. Particles will then be created at the 
boundaries separating inactive domains by (\ref{reac_4}), but
again they will disappear immediately by reaction (\ref{reac_2}).
Hence, when looked at the time scale of the
slow processes (\ref{reac_3}) and (\ref{reac_4}) the dynamics
can be limited to the set of configurations that contain
only empty sites.
(In this limit the creation of two particles on nearest neighbour sites occurs 
with even smaller probability and the effect of (\ref{reac_1}) can therefore 
be neglected). As an example consider:
\begin{equation}
\ldots 
\emptyset_k \emptyset_k \emptyset_k \emptyset_l \emptyset_l 
\ldots \stackrel{(4)}{\rightarrow} \ldots 
\emptyset_k \emptyset_k A \emptyset_l \emptyset_l 
\ldots \stackrel{(2)}{\rightarrow} \ldots 
\emptyset_k \emptyset_k \emptyset_l \emptyset_l \emptyset_l 
\ldots
\end{equation}
The whole process consists of moving the domain wall one unit to the left
with rate (on the time scale of the slow process) equal to $1/2$.
Going in this way through all possibilities one can derive an effective 
dynamics on the slow time scale, which turns out to involve
diffusion, annihilation and coagulation of domain 
walls.

The above heuristic reasoning can be made mathematically
rigorous using a fast rate expansion introduced in \cite{schutz}.
We therefore write  $H$ as
\begin{equation}
H=H_{0}+\mu H_{1}\label{h=h0+h1}
\end{equation}
where $H_{0}$ contains the reactions (\ref{reac_1}),(\ref{reac_3})
and (\ref{reac_4})
while $H_{1}
$ contains only the reactions (\ref{reac_2}). 
In the limit $\mu \to \infty$, it is appropriate to apply the
Schwinger-Dyson formula to the operator $e^{-Ht}$ which 
appears in the formal solution of the master equation (\ref{f1}). One has
\begin{eqnarray}
e^{-(H_{0}+ \mu H_{1})t} &=& e^{-\mu H_{1}t}\Big[
1 - \int_{0}^{t} d\tau_{1} H_{0}(\tau_{1}) \nonumber \\ & &+ \int_{0}^{t} d\tau_{1}
\int_{0}^{\tau_{1}} d\tau_{2} H_{0}(\tau_{1}) H_{0}(\tau_{2}) 
 + \ldots \Big]
\label{f2}
\end{eqnarray}
where
\begin{eqnarray}
H_{0}(\tau) = e^{\mu H_{1}\tau}H_{0}e^{-\mu H_{1}\tau}
\label{f3}
\end{eqnarray}
Using this expansion it can be shown \cite{schutz} that for $\mu \to \infty$ 
the time evolution operator $e^{-Ht}$ becomes
\begin{eqnarray}
\lim_{\mu \to \infty} e^{-(H_{0}+\mu H_{1})t} = 
e^{-{\widetilde H}_{0}t} T^{\star}
\label{f4}
\end{eqnarray}
where
\begin{eqnarray}
{\widetilde H}_{0}= T^{\star} H_{0} T^{\star}
\label{f5}
\end{eqnarray}
and $T^{\star}$ is the projector on the ground states of $H_{1}$, i.e.
\begin{eqnarray}
T^{\star} = \lim_{t \to \infty} e^{-H_{1}t}
\label{f6}
\end{eqnarray}
Equation (\ref{f4}) shows that for $\mu \to \infty$ the generator of the 
effective dynamics is ${\widetilde H}_{0}$ which is nothing but $H_{0}$ 
projected onto the ground states of $H_{1}$. 
In the GCP, the ground states of $H_{1}$ for a chain of length $L$
are the $n^{L}$ configurations without particles and $\widetilde{H}_{0}$ 
is then the effective Hamiltonian of the slow processes projected
in this reduced space. This is the mathematical description 
of the physical arguments given in the beginning of this section.
If one works out the matrix elements of $\widetilde{H}_{0}$ (see the appendix) 
one obtains that the following processes can occur with the indicated rates
\begin{eqnarray}
\emptyset_{k} \emptyset_{l} \emptyset_{k} &\to& \emptyset_{k} 
\emptyset_{k} \emptyset_{k} \ \ \ \mbox{rate 2} \label{18} \\
\emptyset_{k} \emptyset_{l} \emptyset_{l} &\to& \emptyset_{k} 
\emptyset_{k} \emptyset_{l} \ \ \ \mbox{rate 1/2} \nonumber \\
&\to& \emptyset_{k} 
\emptyset_{l} \emptyset_{l} \ \ \ \mbox{rate 1/2} \label{19} \\
\emptyset_{k} \emptyset_{l} \emptyset_{m} &\to& \emptyset_{k} 
\emptyset_{k} \emptyset_{m} \ \ \ \mbox{rate 1} \nonumber \\
&\to& \emptyset_{k} 
\emptyset_{m} \emptyset_{m} \ \ \ \mbox{rate 1} \label{20} 
\end{eqnarray}

It is now useful to interpret the $n$ empty states as the possible spin values 
of an $n$-state Potts model. 
In this language, the processes (\ref{18}-\ref{20}) can be summarized
as follows: the central spin assumes the value of one of its
neighbour spins with equal probability. The dynamics
of our model in the limit $\mu \to \infty$ is therefore
consistent with the requirements of detailed balance for
an $n$-state Potts model at zero temperature. It is generally
expected that if such a dynamics includes a domain wall
diffusion, that then it is critical with a dynamic exponent
$z=2$ \cite{droz}, independently of $n$.
In fact, the exponent $z=2$ can be derived exactly for the case that 
the rate of the process (\ref{18}) equals $1$, and those
of the processes in (\ref{20}) equal 1/2 \cite{derrida}.
Hence, it is quite possible that also for our model, $z=2$ exactly.
For $n=2$ we thus recover the known dynamical exponent $z=2$
in the inactive phase of a model with a PC transition.
For $n \geq 3$ our numerical data strongly suggest that
for $\mu \to \infty$ the GCP is critical, and
hence the exponent $z=2$ must correspond to the dynamical
exponent at criticality for these models. This estimate
is indeed consistent with the value that was determined
numerically for $n=3$ in the previous section.

The result that numerically the exponent $\beta$ is the same for $n=3$ and 
$n=4$, combined with the fact that $z=2$ if $n \geq 3$ leads us to conjecture
that for all $n \geq 3$ the critical exponents are the same, i.e. $\beta=1$, 
$z=2$ and $\nu_{\parallel}=2$.
In the next section we will give further arguments that support this 
conjecture.

It is also possible to study the effective dynamics of the 
GCP for $\lambda \to \infty$. In this limit, the dynamics
of the model when considered on the time scale of the
slow processes (\ref{reac_2}-\ref{reac_4}), will be restricted
to the space of configurations  without particle
pairs.
Each particle present in the system then separates two domains
of empty sites. Therefore particles can be labelled by two indices
and in this way they can be divided into classes. We
will denote by $A_{ij}$ a particle $A$ that separates an 
$\emptyset_{i}$-domain on its left side from an $\emptyset_{j}$-domain
on its right side. In the limit $\lambda \to \infty$, it is useful
to look at the dynamics of these particles.
When $n=1$ there is only one type of particle, and from a determination
of the effective Hamiltonian $\widetilde{H}_{0}$ (see the appendix) we
find that this particle can diffuse, and undergo the reactions
$2A \to A$ and $A \to \emptyset$.
Since there are no processes that create particles, we arrive at the conclusion
that independently of $\mu$, the GCP with $n=1$ must always be in the 
inactive state when $\lambda \to \infty$. 
This is in agreement with our numerical results (see Fig. \ref{FIG01}).

For $n>1$, the situation is less clear. Now there are processes
that both destroy and create particles present in the effective
dynamics (see the appendix). It is therefore in principle possible
to have both an active and an inactive phase, depending on the value
of $\mu$. At this moment, we can draw no firm conclusions
for the form of the phase diagram when $\lambda \to \infty$. On the
basis of our numerical work and on the basis of our results
for $\mu \to \infty$, we believe that the model is probably always
active along that line, at least when $n > 2$.

\begin{figure}[b]
\centerline{
\psfig{file=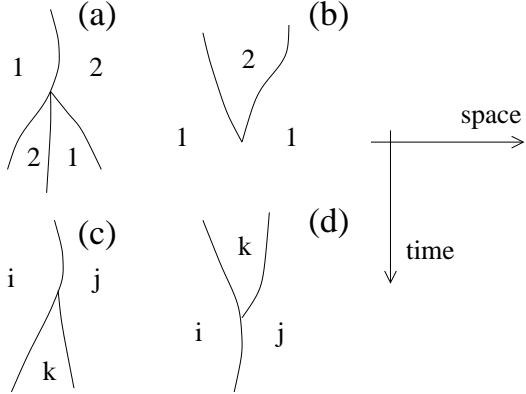,height=5.3cm}}
\vskip 0.2truecm 
\caption{Possible reactions in the coarse-grained representation of the
GCP for $n=2$ (a-b) and $n >2$ (a-b-c-d).
}
\label{FIG07}
\end{figure}

\section{Relation with branching and annihilating walks}
\label{sec:conjecture}

In his paper, Hinrichsen \cite{Hinr97} gave an heuristic argument which relates
the GCP for $n=2$ to the BARW with two offsprings, thus explaining the 
PC universality found numerically. This argument works as follows: indicating
with $X_{ij}$ the domain wall between two configurations $\emptyset_i$ and
$\emptyset_j$ he considered an effective dynamics for the variables $X_{ij}$. 
As seen in the previous section such representation of the original model
becomes exact either in the limit $\lambda \to \infty$ and then $X_{ij}$
coincides with the particle $A_{ij}$, or in the limit $\mu \to \infty$
where $X_{ij}$ coincides with the bond variable $\emptyset_i \emptyset_j$. 
For finite values of these parameters one can still apply this reasoning at a
coarse-grained level. In this case $X_{ij}$ is not a sharp domain wall, but 
an object with a fluctuating thickness. Hinrichsen argued that in this 
coarse-grained representation the most likely reactions for $n=2$ are:
\begin{eqnarray}
X_{ij} \stackrel{(a)}{\rightarrow} X_{ij} X_{ji} X_{ij}
\ \ \ \ \ \ \ 
X_{ij} X_{ji} \stackrel{(b)}{\rightarrow} 0
\label{barwe}
\end{eqnarray}
(for $n=2$, one has obviously $i=1$, $j=2$ or viceversa). An example 
of such reactions is shown in Fig. \ref{FIG07} (a) and (b). Notice that the 
reactions (a) and (b) given in (\ref{barwe}) are those for a 
branching-annihilating random walk (BARW) with two offsprings, which 
suggests indeed, as found numerically that the universality class is PC. 

These arguments can be extended to the case $n > 2$, where there is still
the possibility of having reactions of type (a) and (b), but also reactions
involving three different domains ($i\neq j$, $i \neq k$ and $j \neq k$).:
\begin{eqnarray}
X_{ij} \stackrel{(c)}{\rightarrow} X_{ik} X_{kj}
\ \ \ \ \ \ \ 
X_{ik} X_{kj} \stackrel{(d)}{\rightarrow} X_{ij}
\label{barwngt2}
\end{eqnarray}
When $n > 2$ there are actually $n(n-1)/2$ domain walls, and the model
described by reactions (\ref{barwe}) and (\ref{barwngt2}) is now a BARW with
more than one type of particles. 
To our knowledge this type of model has not been studied yet, but we expect it 
to be in the same universality class as the GCP with $n > 2$, i.e. always
in an active state except when the rates for the processes (a) and (c) are 
zero, and with exponents $z=2$, $\beta=1$ and $\nu_\|=2$.

There is a BARW with more than one type of particles that has attracted some 
attention recently. The model was introduced by Cardy and T\"auber 
\cite{Card98}, who considered a system with $N$ different particles 
$A^\alpha$, where $\alpha=1$, $2$, \ldots $N$ which diffuse and
undergo the reactions:
\begin{eqnarray}
2 A^\alpha \rightarrow  0 
&&
\ \ \ \ \ \ \ \
\mathrm{with\ rate\ 1}\label{cardy_1}\\
A^\alpha \rightarrow A^\alpha A^\alpha A^\alpha
&&
\ \ \ \ \ \ \ \
\mathrm{with\ rate\ \sigma}\label{cardy_2}\\
A^\alpha \rightarrow A^\alpha A^\beta A^\beta 
&&
\ \ \ \ \ \ \ \
\mathrm{with\ rate\ \sigma^\prime/(N-1)}\label{cardy_3}
\end{eqnarray}             
where in the last reaction it is understood that $\alpha \neq \beta$.
The {\it bosonic} version of this model has precisely the same exponents we 
determined for the GCP with $n>2$ \cite{Card98}.
Notice that the coarse-grained representation of the GCP as defined by the 
reactions in Fig. \ref{FIG07} and the model defined by the reactions 
(\ref{cardy_1},\ref{cardy_2},\ref{cardy_3}) do not actually coincide. While 
there is an obvious correspondence between
(\ref{cardy_1}), (\ref{cardy_2}) with (b), (a) of Fig. \ref{FIG07},
the reaction (\ref{cardy_3}) does not have any obvious counterpart. It is not a
priori clear therefore that the two models are in the same universality class.
The coincidence of the critical exponents therefore suggests that one could replace 
(\ref{cardy_3}) with other reactions, as for instance $A^\alpha \to A^\gamma 
A^\delta$, without changing the universality class. To our knowlegde, BARW
models with this kind of reactions have not been studied yet. They form an 
interesting subject for further investigation.

The model of Cardy and T\"auber has raised some interest recently since it 
has been found that fermionic and bosonic versions of the model are in 
different universality classes \cite{Kwon00}. In the fermionic version only 
one particle per lattice site is allowed, which implies that particles of 
different species block each other. 
In the fermionic version of the model it makes for instance a difference wether
the two offsprings produced by the reaction (\ref{cardy_3}) are placed to the 
same side or at opposite sides of the parent particle \cite{Kwon00,Odor00}.
If, for instance, they are placed at opposite sides the offsprings cannot 
annihilate through the reaction (\ref{cardy_1}) because of the presence of the
parent particle that blocks them. In the bosonic model the two offspring can
instead always recombine.

It is important to stress here that in the multispecies BARW model we 
constructed ((\ref{barwe})-(\ref{barwngt2})) from a coarse-grained 
representation of the GCP there are no blocking effects.
By the very construction of the model two domain walls approaching each other
can always annihilate. Therefore even if the model is clearly of fermionic 
nature its universality class, as found numerically, is that of the bosonic 
multispecies BARW.

\section{The effect of breaking the symmetry}

\label{sec:breaking}

As a final point we consider the effect of breaking the permutation symmetry
of the inactive states of the model. For $n=2$ Hinrichsen \cite{Hinr97} 
explicitely broke the $Z_{2}$ symmetry by choosing $\mu_{1}\neq\mu_{2}$ in 
the reaction (\ref{reac_2}). As a result he found the system to switch from 
PC to DP behavior, which was understood as PC being related to the presence 
of an exact $Z_{2}$ symmetry.

\begin{figure}[b]
\centerline{
\psfig{file=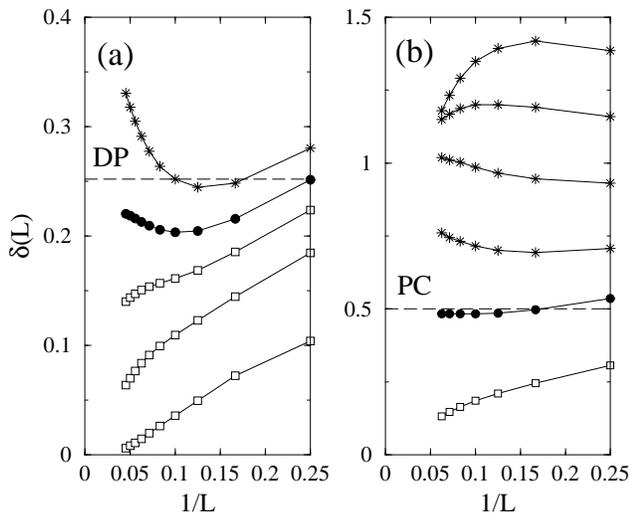,height=7cm}}
\vskip 0.2truecm 
\caption{Plot of $\protect\delta$ for the GCP with $n=3$ and (a) $
\protect\mu_1 /2 = \protect\mu_2 = \protect\mu_3$, (b) $2 \protect\mu_1 = 
\protect\mu_2 = \protect\mu_3$. Both cases were calculated with $\protect
\lambda = 0.5$ and $p^{\prime}=1.5$. Symbols refer to curves calculated in the
inactive phase (stars), at the critical point (filled circles) and in the
active phase (empty squares). At criticality $\delta(L)$ converges towards
the values of $\beta/\nu_\perp$ expected for DP (a) and PC (b), and indicated
by horizontal dashed lines in the figure.
Notice the two distinct behaviors of $\protect\delta$ in the inactive phase.
}
\label{FIG08}
\end{figure}

For $n=3$, we now perform a similar symmetry-breaking by considering the
following two cases: (a) $\mu_{1}/2=\mu_{2}=\mu_{3}$ and (b) $2\mu_{1}=\mu
_{2}=\mu_{3}$. In both cases the system has a $Z_{2}$ symmetry due to the
equivalence of the states $\emptyset_{2}$ and $\emptyset_{3}$.
The difference is that starting from a random configuration in the 
first
case the system is more likely to reach the absorbing states
$(\emptyset_{2},\emptyset_{2},\emptyset_{2},\ldots)$ and 
$(\emptyset_{3},\emptyset_{3},\emptyset_{3},\ldots)$ compared with
$(\emptyset_{1},\emptyset_{1},\emptyset_{1},\ldots)$, while in
the second case the reverse is true.
We calculated
the particle density $\rho_{L}(\lambda,\mu_{1},\mu_{2},\mu_{3})$ as before,
using the boundary-term (\ref{inject}). Depending on the phase in which the 
model is, the density will behave as $\rho_{L}=\rho_{0}+Ce^{-aL}$ (in the 
active region), $\rho_{L} \sim L^{-\beta/\nu_{\perp}}$ (at criticality), or
$\rho_{L} \sim e^{-aL}$ (in the inactive phase). If we define $\delta
(L)=-\ln\left[  \rho_{L+1}/\rho_{L}\right]  /\ln\left[  (L+1)/\ln
(L-1)\right]  $, we expect $\lim_{L\rightarrow\infty}\delta(L)$ to be zero in
the active phase, to be $\beta/\nu_{\perp}$ at critical points, and
$+\infty$ in the inactive phase. In Fig. \ref{FIG08} we plotted $\delta(L)$ 
as a function of $1/L$ for cases (a) and (b) with the choice $\lambda=0.5$ and
injection rate $p^{\prime}=1.5$. In both cases for small $\mu$ one finds
the typical scaling behavior of the active phase with $\delta(L)\rightarrow0$
just as for the case $\mu_{1}=\mu_{2}=\mu_{3}$. For large $\mu_{k}$ however
the situation differs from the symmetric model. For (a), we find $\delta
(L)\rightarrow+\infty$ for large $\mu$, i.e. one has a standard inactive phase
with a particle density exponentially small in $L$. In case (b),
$\delta(L)$ becomes equal to $1$, implying that the inactive phase is itself
critical with $\beta/\nu_{\perp}=1$. In between the active and the inactive
phase we have a critical point where $\delta(L)$ is going to a distinct finite 
value. The critical point estimates of $\delta(L)$ are marked by filled 
circles in Fig. \ref{FIG08}. For case (a) the critical point is at 
$\mu_{1}\approx0.64$ , while for (b) it is at $\mu_{1}\approx0.42$. The value 
of $\beta/\nu_{\perp}$ agrees with that of DP and PC respectively, as can be 
seen in Fig. \ref{FIG08} where the critical values of these universality 
classes are indicated with a dotted line.

This indicates that on breaking the symmetry of the inactive states the
remaining symmetry of the \textit{dominant} rates $\mu_{i}$ determines
the critical behavior.
In case (b) $\mu_{1}<\mu_{2}=\mu_{3}$ the dominant rates still have a $Z_{2}$ 
symmetry leading to the PC universality class, while in case (a) $\mu_{1} >
\mu_{2}=\mu_{3}$, DP critical behavior is recovered.

\section{Conclusions}
\label{sec:concl}

In this paper, we studied a generalised contact process first introduced by 
Hinrichsen. The major part of our results were obtained by applying the DMRG 
to the model. With this technique we verified that for $n=2$, the critical 
line of the model is in the PC universality class, consistent with earlier 
results coming from simulations.

A first set of new results was obtained for the case $n=3$, for which we
found that the model is always active, except when $\mu \to \infty$,
which corresponds to the critical line of the model. From our
numerical work we determined the critical exponents to be
equal to $z=2, \nu_{\parallel}=2 $ and $\beta=1$. Using well
established scaling laws \cite{Gras79}, other exponents
can be determined from these three. For $n=4$, we found evidence
that the phase diagram is the same, and that the critical exponent 
$\beta$ also equals 1.
Secondly, using a fast rate expansion that becomes exact for $\mu \to
\infty$, we were able to argue that in that limit $z=2$.
It can be hoped that by examining the model for $\mu^{-1}$ small
using perturbation techniques, it may be possible to determine
also the exponents $\beta$ and $\nu_{\parallel}$ exactly.
On the basis of these numerical and exact results we conjectured
that the universality class of the model is the same for all
$n \geq 3$.

The exponent values that we found for $n \geq 3$ coincide with
those of the BARW model with more than one type of particles
introduced by Cardy and T\"auber \cite{Card98}. 
We were able to give an heuristic argument that explains why the two models 
could be in the same universality class. It is interesting to remark
that despite many attempts the number of universality classes
found for phase transitions out of an adsorbing phase, remains
very limited. It could have been hoped {\it \`{a} priori} that for
the generalised GCP studied here, new universality classes
could appear for $n>2$. In a sense our results show that this
is true, but only in the least exciting way possible: the
universality class does not depend on $n$, and moreover the
exponents take on rather trivial values.
One could hope that by lowering the permutation symmetry to a 
$Z_{n}$-symmetry, other universality classes could appear for $n \geq 4$. 
This could be done, e.g. by having the rates of the process (\ref{reac_4})
depend on $|k-l|\ \mbox{mod}(n)$. But since this would only make a difference 
for $n \geq 4$, it will probably be difficult to investigate such a model with 
the numerical techniques currently available.

We also verified that if one breaks the permutation symmetry of the model with 
$n=3$, one recovers a DP or PC universality, suggesting that it is the symmetry
of the largest rates that determines the universality class.

Finally, we remark that the consistency of the DMRG results with those coming 
from simulation for $n=2$, or with the exactly determined value of $z$ for $n 
\geq 3$, shows convincingly that the DMRG can be trusted as a powerful method 
in the study of (criticality in) non equilibrium systems.

\acknowledgements 
JH and CV thank the IUAP, Belgium for financial support. EC is grateful to
INFM for financial support through PAIS 1999.             

\section*{Appendix}
In this appendix we will calculate explicitly $
\lim_{r\to\infty}e^{-(H_{0}+rH_{1})t}$ where $r$ is one of the
rates of the GCP. In section III
 it was shown that this reduces to the calculation of $\tilde{H}
_{0}=T^{\ast}H_{0}T^{\ast}$, the Hamiltion that describes the effective dynamics.
Since $T^{\ast
}=\lim_{t\to\infty}e^{-H_{1}t}$ this is a projection of $H_{0}$ on the
ground states of $H_{1}$.

Let us denote the right ground states of $H_{1}$ as $\left|
\psi_{i}\right\rangle $ and the left ones as $\left\langle p_{i}\right|  $.
\begin{eqnarray}
H_{1}\left|  \psi_{i}\right\rangle  & =0\\
\left\langle p_{i}\right|  H_{1}  & =0
\end{eqnarray}
The physical meaning of $\left|  \psi_{i}\right\rangle $ is evident: they are
the stationary states of $H_{1}$, any state $\left|  \psi\right\rangle $ will
under the dynamics of $H_{1}$ relax into one of these $\left|  \psi
_{i}\right\rangle $. The left ground states $\left\langle p_{i}\right|  $ can
be interpreted as linear functionals giving the corresponding probabilities:
any state $\left|  \psi\right\rangle $ will under the dynamics of $H_{1}$
relax into the ground state $\left|  \psi_{i}\right\rangle $ with probability
$\left\langle p_{i}\right.  \left|  \psi\right\rangle $ (where we assumed
that $\left\langle p_{i}\right|  $ are normalized: $\langle p_{i}
|  \psi_{j}\rangle =\delta_{ij}$). Using this notation we can write
the projection operator $T^{\ast}$ as
\begin{equation}
T^{\ast}=\sum_{i}\left|  \psi_{i}\right\rangle \left\langle p_{i}\right|
\end{equation}
Since clearly $\left[  \sum_{i}\left\langle p_{i}\right|  \right]  \left|
\psi\right\rangle =1$ for any (normalized) state $\left|  \psi\right\rangle $,
this projection conserves probability, meaning that when $H_{0}$ is a
stochastic operator, so is $\tilde{H}_{0}$. Furthermore we can write the
transition rates of the effective dynamics between state $\left|  \psi
_{i}\right\rangle $ and $\left|  \psi_{j}\right\rangle $ as
\begin{equation}
\text{rate}\left(  \left|  \psi_{i}\right\rangle \rightarrow\left|  \psi
_{j}\right\rangle \right)  =\left\langle p_{j}\right|  (-H_{0})\left|  \psi
_{i}\right\rangle \label{transitionrates}
\end{equation}
These matrix elements determine the effective dynamics, and we will now
calculate them explicitly.

We first study the limit  where the rate $\mu$ of the processes 
$A\emptyset_{k},\emptyset_{k}A\rightarrow\emptyset_{k}
\emptyset_{k}$ goes to infinity. The Hamiltonian is of the form:
\begin{equation}
H=H_{0}+\mu H_{1}
\end{equation}
where $H_{0}$ is the generator of the processes (\ref{reac_1}),
(\ref{reac_3}), (\ref{reac_4})  and
$H_{1}$ is the generator of process (\ref{reac_2}) with the factor $\mu$
brought out. In this case the ground states $\left|  \psi_{i}\right\rangle $ of
$H_{1}$ are all configurations containing no particles $A$. Firstly, we
note that the processes (\ref{reac_1}) and (\ref{reac_3}) can only 
act on
configurations containing particles, so they cannot contribute to the rates
(\ref{transitionrates}) of the effective dynamics and we redefine $H_{0}$
without them.

When we now project $H_{0}$ which contains only 2-site interactions, onto the
$\left|  \psi_{i}\right\rangle $, the resulting operator will contain 3-site
interactions. It is therefore convenient to first rewrite $H_{0}$ as a 3-site
operator. Since we only have reaction (\ref{reac_4}) left in $H_{0}$, this
becomes ($k\neq l\neq m$)
\begin{equation}
\begin{tabular}
[c]{ll}
$\emptyset_{k}\emptyset_{l}\emptyset_{k}\longrightarrow\emptyset_{k}
A\emptyset_{k}$ & rate 2\\
$\emptyset_{k}\emptyset_{l}\emptyset_{m}\longrightarrow\emptyset_{k}
A\emptyset_{m}$ & rate 2\\
$\emptyset_{k}\emptyset_{l}\emptyset_{l}\longrightarrow\emptyset_{k}
A\emptyset_{l}$ & rate 1\\
$\emptyset_{k}\emptyset_{l}A\longrightarrow\emptyset_{k}AA$ & rate 1
\end{tabular}
\label{three-site}
\end{equation}
(together with some reactions that are obtained by reflection). We
finally notice that the last process of (\ref{three-site}) is again not
relevant for the projection on $\left|  \psi_{i}\right\rangle $, and determine
the effective dynamics in the following diagram
\begin{eqnarray}
\begin{tabular}
[c]{ccc}
\begin{tabular}
[c]{c}
reaction\\
\multicolumn{1}{l}{with rate...}
\end{tabular}
&
\begin{tabular}
[c]{c}
projection\\
\multicolumn{1}{l}{with probability...}
\end{tabular}
& netto rate\\
&  & \\
$\emptyset_{k}\emptyset_{l}\emptyset_{k}\stackrel{\text{rate 2}}
{\longrightarrow}\emptyset_{k}A\emptyset_{k}$ & $\stackrel{\text{1}
}{\longrightarrow}\emptyset_{k}\emptyset_{k}\emptyset_{k}$ & rate $2.1=2$\\
$\emptyset_{k}\emptyset_{l}\emptyset_{m}\stackrel{\text{rate 2}}
{\longrightarrow}\emptyset_{k}A\emptyset_{m}$ & $\stackrel{\text{1/2}
}{\longrightarrow}\emptyset_{k}\emptyset_{k}\emptyset_{m}$ & rate $2.\frac
{1}{2}=1$\\
& $\stackrel{\text{1/2}}{\longrightarrow}\emptyset_{k}\emptyset_{m}
\emptyset_{m}$ & rate $2.\frac{1}{2}=1$\\
$\emptyset_{k}\emptyset_{l}\emptyset_{l}\stackrel{\text{rate 1}}
{\longrightarrow}\emptyset_{k}A\emptyset_{l}$ & $\stackrel{\text{1/2}
}{\longrightarrow}\emptyset_{k}\emptyset_{k}\emptyset_{l}$ & rate $1.\frac
{1}{2}=\frac{1}{2}$\\
& $\stackrel{\text{1/2}}{\longrightarrow}\emptyset_{k}\emptyset_{l}
\emptyset_{l}$ & rate $1.\frac{1}{2}=\frac{1}{2}$
\end{tabular}
\end{eqnarray}
which are the processes and their corresponding rates already given in
(\ref{18}-\ref{20}). Note that this
calculation is exactly the same for $\mu=\lambda\rightarrow\infty$.

Next, we consider now the limit $\lambda \to \infty$. In this case we have
\begin{equation}
H=H_{0}+\lambda H_{1}
\end{equation}
where $H_{0}$ is the generator of the processes 
(\ref{reac_2}-\ref{reac_4}), and
$H_{1}$ is the generator of process (\ref{reac_1}) with the factor $\lambda$
brought out. The ground states of $H_{1}$ are now all configurations containing
no particle pairs. In contrast with the previous case, all processes of
$H_{0}$ are now relevant for the projection on the ground states of $H_{1}$.

We will start with the case $n=1$, where there is only one inactive state
$\emptyset$, and process (\ref{reac_4}) can a priori not take place. For
process (\ref{reac_3}) we again use the 3-site representation, while process
(\ref{reac_2}) is so simple that we keep the 2-site representation. We then
get the following effective dynamics
\begin{equation}
\begin{tabular}
[c]{ccc}
\begin{tabular}
[c]{c}
reaction\\
\multicolumn{1}{l}{with rate...}
\end{tabular}
&
\begin{tabular}
[c]{c}
projection\\
\multicolumn{1}{l}{with probability...}
\end{tabular}
& netto rate\\
&  & \\
$A\emptyset\stackrel{\text{rate }\mu}{\longrightarrow}\emptyset\emptyset$ &
$\stackrel{\text{1}}{\longrightarrow}\emptyset\emptyset$ & rate $\mu.1=\mu$\\
$A\emptyset\emptyset\stackrel{\text{rate 1}}{\longrightarrow}AA\emptyset$ &
$\stackrel{\text{1/2}}{\longrightarrow}A\emptyset\emptyset$ & rate $1.\frac
{1}{2}=\frac{1}{2}$\\
& $\stackrel{\text{1/2}}{\longrightarrow}\emptyset A\emptyset$ & rate
$1.\frac{1}{2}=\frac{1}{2}$\\
$A\emptyset A\stackrel{\text{rate 2}}{\longrightarrow}AAA$ & $\stackrel
{\text{1/2}}{\longrightarrow}A\emptyset A$ & rate $2.\frac{1}{2}=1$\\
& $\stackrel{\text{1/4}}{\longrightarrow}\emptyset A\emptyset$ & rate
$2.\frac{1}{4}=\frac{1}{2}$\\
& $\stackrel{\text{1/8}}{\longrightarrow}A\emptyset\emptyset$ & rate
$2.\frac{1}{8}=\frac{1}{4}$\\
& $\stackrel{\text{1/8}}{\longrightarrow}\emptyset\emptyset A$ & rate
$2.\frac{1}{8}=\frac{1}{4}$
\end{tabular}
\label{zwappa}
\end{equation}
For $\lambda\rightarrow\infty$ and $n=1$ we find the effective dynamics to
contain only diffusion and destruction of particles. Because of the first
reaction appearing in (\ref{zwappa}),
the decay of particles is exponentially fast, meaning that for any
finite value of $\mu$ this system is non-critical.

For the case $n>1$ also process (\ref{reac_4}) has to be taken into account. As
a consequence, two different neighbouring inactive domains remain active.
For example
\begin{eqnarray}
\emptyset_{k}\emptyset_{l}\emptyset_{l}\longrightarrow\emptyset_{k}
A\emptyset_{l}
\label{ad}
\end{eqnarray}
remains a process of the effective dynamics. One can easily
construct the complete effective dynamics, but this does not lead
to much further insight in the phase diagram. One can only conclude
that because of the presence of the process (\ref{ad}), it is
in principle possible that both active and inactive phases are
present.


\end{multicols}


\begin{thebibliography}{99}
\bibitem{nota} The transformation (\ref{shift}) was also implemented in Ref.
\cite{2a3a}.
\bibitem{nota1} It should be noted that rigorously for the calculation of
the exponent $z$ one should first extrapolate the value of $\mu_{\rm max}(L)$
to the thermodynamic limit $L \to \infty$ to obtain an estimate of the critical
value $\mu_c$. Then perform another extrapolation $Y_L (\mu_c) \to -z$. 
However in the present model both numerical procedures give the same 
extrapolated value for $z$. For a more detailed discussion on this point see 
the Appendix B of Ref. \cite {2a3a}.
Notice also that since the maxima of $Y_L(\lambda,\mu)$ are not so pronounced 
one can locate accurately their ordinates, while their abscissas are somewhat 
less accurate. This is the reason why extrapolated value of $z$ are more 
accurate than the location of the critical point itself.
\end{thebibliography}
\end{document}